# Depth-resolved Analysis of Metamagnetic Phase Transition of FeRh Alloy in Composite Multiferroic FeRh/BaTiO$_3$

*Attila Lengyel[1], Gábor Bazsó[1], Aleksandr I. Chumakov[2], Dénes L. Nagy[1], Gergő Hegedűs[1], Dimitrios Bessas[2], Zsolt E. Horváth[3], Norbert M. Nemes[4], Maria A. Gracheva[5], Edit Szilágyi[1], Szilárd Sajti[1], Dániel G. Merkel[1, 3]*

[1] Wigner Research Centre for Physics, P.O.B. 49, 1525 Budapest, Hungary

[2] ESRF - The European Synchrotron Radiation Facility, 71, Avenue des Martyrs, 38043 Grenoble, France

[3] Centre for Energy Research, P.O.B. 49, 1525 Budapest, Hungary

[4] GFMC, Departamento de Física de Materiales, Universidad Complutense de Madrid, E-28040 Madrid, Spain

[5] Institute of Chemistry, Eötvös Loránd University, Pázmány Péter sétány 1/A, 1117 Budapest, Hungary

*email of corresponding author:* lengyel.attila@wigner.hu

## Abstract

We report on the depth dependence and technological limits of the metamagnetic phase transition of the iron rhodium alloy as a function of temperature, external magnetic and electric fields in the thin-film FeRh/BaTiO$_3$ multiferroic determined by grazing-incidence nuclear resonance scattering measurements. The change of temperature induces a continuous and homogeneous antiferromagnetic / ferromagnetic phase transition through the entire FeRh layer, except in the near-substrate region. Application of external magnetic field does not affect this mechanism in contrast to electric field that changes it fundamentally via piezoelectric strain: the phase transition of the alloy propagates from the substrate up to a height defined by the combination of temperature and external magnetic field, as soon as the applied electric field reaches a temperature-determined voltage limit.

## 1. Introduction

In 2020, the global electricity consumption was 22.5 ×10$^3$ TWh/year, a figure forecasted to increase by 2030 to 27.5 × 10$^3$ TWh/year and 40 × 10$^3$ TWh/year at best and at worst, respectively [1, 2]. By the same year, the electricity footprint of communication technology alone can reach about 8 ×10$^3$ TWh/year [3]. Ignoring such growth may have a serious impact on the environment, however these dramatic increases could be curbed by the use of new technologies and materials which help to optimize the performance of electronic devices [4]. Composite multiferroics, which are composed of two materials of different ferroic properties [5, 6], are among the novel materials that can be used to increase the efficiency of devices based on the principles of spintronics [7, 8, 9], magnetic switching [10, 11, 12], magnetic refrigeration [13, 14, 15, 16], biomechanical energy harvesting [17, 18, 19], giant magnetoresistance [20, 21, 22] and photovoltaics [23, 24, 25].

The iron rhodium alloy is one of the most promising examples of composite multiferroics, due to its technologically exploitable mechanical and magnetic properties [26]. The equiatomic FeRh alloy has a temperature-induced, fully reversible, *antiferromagnetic* (AFM) ↔ *ferromagnetic* (FM) phase transition in the operating temperature range of modern electronic devices [27, 28, 29]. At room temperature, the equiatomic FeRh is in the AFM phase with a CsCl-type bcc-based B2 crystal structure [30] (see also Fig. 1 of Ref. [31]). In addition to the changing magnetic order, the transition of FeRh has effects on other properties of the alloy, as well. First, it involves the change of the magnetic moments of iron and rhodium atoms from the room-temperature AFM values of 3.2 $\mu_B$ (Fe) and 0.0 $\mu_B$ (Rh), to the high-temperature FM values of 3.1 $\mu_B$ (Fe) and 0.9 $\mu_B$ (Rh) [32]. Second, the phase transition directly influences the electrical resistivity of FeRh [33]. Third, the FM → AFM phase transition is accompanied with the decrease of lattice parameter by about 1 % [34, 35]. By reversing the third phenomenon, mechanical compression can be used to trigger the FM → AFM magnetic phase transition of the alloy [36, 37, 38].

To create a multiferroic composite, FeRh thin films are often coupled with piezoelectric BaTiO$_3$ (BTO) ceramic [26, 38], the epitaxial coupling of which allows the alternation of the phase composition of the alloy (along with its above listed consequences) with external electric field via piezoelectric strain [26, 39, 40, 41, 42].

Notwithstanding the detailed information on its exceptional properties, the optimal utilization of the FeRh/BTO multiferroic requires the in-depth knowledge of how the phase composition of the FeRh alloy changes as a function of technologically important external parameters (temperature, magnetic and electric fields). The phase transition of the alloy on its surface can be efficiently imaged by either magnetic force [43] or electron microscopy [44, 45], so the horizontal map of its AFM / FM structure has been by now thoroughly explored [46, 47]. In contrast, only a handful studies [48, 49, 50, 51] describing the depth-dependent AFM / FM phase structure of the FeRh layer are available, while no studies were devoted so far to investigating the depth profile of the phase transition as a function of temperature, external magnetic and electric fields. The most detailed description of the phase-selective depth profile of FeRh was given by a *polarized neutron reflectivity* (PNR) study [50], according to which the magnetic structure of the alloy in the FeRh/MgO composite was vertically inhomogeneous at room temperature, with a thin FM layer near the substrate. This FM phase may had been an effect of the epitaxial strain caused by the lattice mismatch between the substrate and the FeRh layer [52, 53, 54], so it could be suspected that the depth profile described in the PNR study varied depending on the substrate used. In a magnetization / ferromagnetic resonance study correlation was established between the strain present in FeRh epitaxial films and their magnetic behavior [31]. Therefore, for the optimal utilization of the FeRh/BTO composite multiferroic, a comprehensive investigation is still needed to reveal the depth profile of the phase transition of the alloy as a function of technologically important external parameters.

In the present work, the FeRh layer was deposited by *molecular beam epitaxy* (MBE) on the BTO substrate. Its atomic composition was specified by *Rutherford backscattering spectrometry* (RBS). The overall crystal structure and lattice parameters of the alloy were determined by *X-ray diffractometry* (XRD), while the thickness of the film was measured by *X-ray reflectometry* (XRR). The magnetic properties of the alloy were calculated from *vibrating-sample magnetometry* (VSM) data and the iron microenvironments were determined by *conversion-electron Mössbauer spectroscopy* (CEMS). Finally, the depth profile of the phase transition of the FeRh was investigated by *grazing-incidence nuclear resonance scattering* (GI-NRS).

Here we report a depth-resolved analysis of the AFM ↔ FM metamagnetic phase transition of the FeRh alloy in the composite multiferroic FeRh/BaTiO$_3$. We find that the FeRh is always in FM phase in the few nm vicinities of both top and bottom edges of the alloy. We show that these two FM sublayers can never be transformed into AFM phase by temperature change. Conversely, anywhere else within the FeRh layer the change of temperature induces a continuous and homogeneous AFM / FM metamagnetic phase transition. This phase structure and the phase transition remain the same, regardless of external magnetic field, however the maximum achievable AFM fraction depends on the strength of the external magnetic field. In contrast, the applied electric field induces the magnetic phase transition with a bottom → up orientation as soon as the voltage reaches a certain limit set by the temperature of the alloy. Finally, we also demonstrate that the layer thickness, where the electric field can induce the phase transition is adjustable by changing the temperature and external magnetic field.

## 2. Experimental

### 2.1. Sample preparation

The BaTiO$_3$(100) substrate (5 mm × 5 mm × 1 mm, rectangular) was purchased from Alineason – Material & Technology GmbH. Before the sample preparation, the substrate was cleaned in ultrasonic ethanol bath, then it was baked out under ultra-high vacuum conditions at 873 K for 1800 seconds. The $^{57}$FeRh alloy of 18.0 nm nominal thickness was deposited on the BTO substrate using the MBE apparatus of the *Wigner Research Centre for Physics* (Wigner RCP). Electron gun at growth rate of 0,0276 Å/s for $^{103}$Rh and effusion cell at growth rate of 0,0154 Å/s for $^{57}$Fe evaporation were used, values corresponding to the equiatomic composition. During deposition, the temperature of the substrate was kept at 903 K while the pressure in the growing chamber never exceeded $3.7 \times 10^{-8}$ mbar. The quality of the epitaxial growth was monitored by in-situ *reflection high-energy electron diffractometry* (RHEED). Later a 20 nm thick gold layer was deposited on the opposite side of the BTO substrate for electric contact.

## 2.2. Characterization methods

### 2.2.1. Rutherford backscattering spectrometry

The RBS measurement of the $^{57}$FeRh layer was performed using 2 MeV $^4$He$^+$ ion beam obtained from the 5 MV Van de Graaff accelerator of the Wigner RCP. The beam was collimated to the necessary dimensions of $0.5 \times 0.5$ mm$^2$ with two sets of four-sector slits. The measurements were performed with an ORTEC ruggedized partially depleted silicon radiation detector of a solid angle of 4.754 msr mounted at a scattering angle of 165° and at tilt angles 7° and 60°. The tilt angle 7° rather than perpendicular incidence was chosen and the sample was continuously rotated during the measurement around the azimuth axis to avoid channeling effects in the substrate. The measurement with 60° tilt angle was necessary since only taking spectra at two different tilt angles assured to make a distinction between the cases that an attenuated signal came from a lower-mass nucleus close to the surface or from a heavier nucleus in a deeper region of the thin film (mass-depth-ambiguity). The dose of the measurement was 4 μC. The ion current typically of 8 nA was measured by a transmission Faraday cup [55]. To reduce the surface contamination, liquid N$_2$ trap was used. The pressure in the scattering chamber was about $2.5 \times 10^{-6}$ mbar during the experiments. The RBS data were evaluated by the RBX code [56].

### 2.2.2. X-ray diffractometry

XRD experiment was carried out at the *Centre for Energy Research* using a D8 Discover (Bruker AXS, Karlsruhe, Germany) diffractometer. Cu Kα radiation (λ = 1.5418 Å) was used for the measurement. To decrease beam divergence and to improve the parallelism of the beam, 1 mm slits were used at the source and the detector. Furthermore, a 90° rotated Soller slit was installed between the sample and the detector-side slit. At last, a secondary monochromator was used at the detector side, to achieve better signal-to-noise ratio. For the evaluation of the XRD results, the Diffrac.EVA [57] program was used.

### 2.2.3. Conversion-electron Mössbauer spectroscopy

CEMS measurement was performed at the Wigner RCP using a conventional WissEl/DMSPCA Mössbauer spectrometer operated in sinusoidal drive mode at 16 Hz drive frequency. The activity of the $^{57}$Co(Rh) single-line Mössbauer source was 621 MBq at the time of the measurement. The resonant conversion electrons were detected with a home-made gas-flow single-wire proportional counter of 1 mm distance between sample and anode wire, working with a mixture of 96 % v/v He and 4 % v/v CH$_4$ gas at bias voltage 884 V. The distance between the source and the sample was 53 mm. Both source and sample were kept at room temperature. The spectrum was evaluated using the MossWinn 4 code [58].

### 2.2.4. Grazing-incidence nuclear resonance scattering and X-ray reflectometry

GI-NRS and XRR experiments were carried out at the Nuclear Resonance beamline [59] ID18 of the *European Synchrotron Radiation Facility* (ESRF). The measurements were performed in 4 bunch mode at 14.414 keV, the energy corresponding to the $1/2 \leftrightarrow 3/2$ nuclear transition of $^{57}$Fe, with a beam of 0.5 meV energy bandpass. The beam was focused by a Kirkpatrick-Baez mirror system both horizontally and vertically to 20 μm and 8. 7 μm, respectively. The sample was mounted in a custom-built vacuum chamber (**Fig. 1.**) that allowed temperature (273 K – 400 K) control and adjustable voltage (0 V – 200 V) and magnetic field (0 mT – 150 mT) to be applied on the sample during the measurements. The GI-NRS quantum-beat patterns and XRR reflectograms were analyzed using the in-house (Wigner RCP) developed FitSuite program [60].

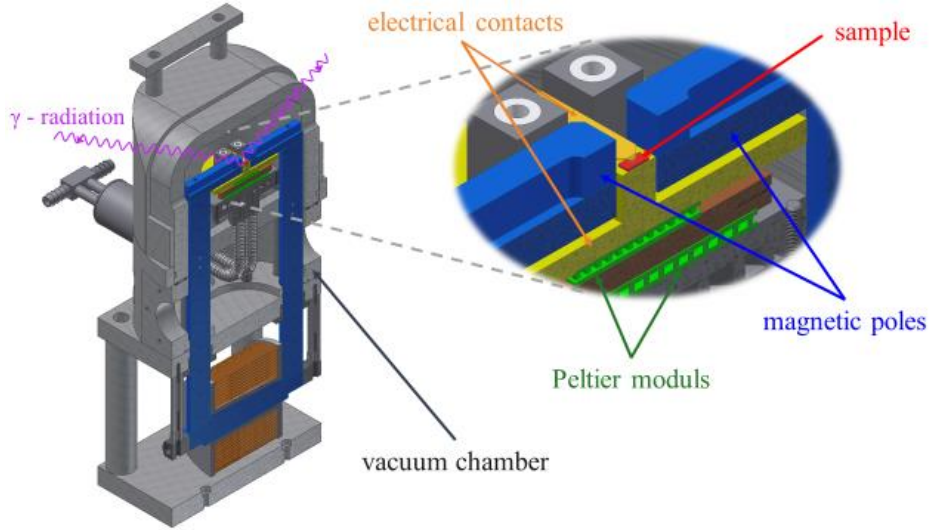

**Fig. 1**. Sketch of the sample holder designed for the GI-NRS experiment

The GI-NRS experiment was performed in four groups of measurements. The 1st group was performed without applying any external electric or magnetic field. The temperature of the sample was raised directly to 394 K, then it was gradually lowered to 299 K, meanwhile GI-NRS quantum-beat patterns were recorded at three different grazing angles (3.49 mrad, 4.19 mrad and 4.71 mrad), at each temperature step (cooling phase). After this, the temperature of the sample was gradually raised back to 394 K, again with 3 – 3 GI-NRS measurements, at each selected grazing angle, at each temperature step (heating phase). The 2nd group of GI-NRS measurements were performed similarly to the 1st ones, but with a 150 mT external magnetic field applied to the sample. During the 3rd group of GI-NRS measurements, the sample was heated up to 394 K; then the voltage in the electric contacts (yellow parts in **Fig. 1.**) was gradually increased up to 100 V (i.e. 0 – 100 kV/m electric field (*E*) was generated in the 1 mm thick BTO substrate), meanwhile 3 – 3 GI-NRS measurements were performed at each selected grazing angle and at each voltage step. After the 100 kV/m measurements, the voltage was switched off, and the sample was cooled down to the next temperature step, then the whole electric field build up process was repeated. Finally, the 4th group of GI-NRS experiments was performed in the same way as the 3rd one, but with 150 mT external magnetic field applied.

### 2.2.5. *Vibrating sample magnetometry*

The magnetic moments were measured with a vibrating sample magnetometer from Quantum Design in a Physical Properties Measurement System equipped with a 9 T superconducting magnet (PPMS-9T) at *Complutense University of Madrid* (UCM). All VSM measurements were performed in five orientations: four times in in-plane orientations [(100), (010), (110) and (-110)] and once in perpendicular direction (001), in the order listed. Magnetic hysteresis loops were measured in 10 K steps in the temperature range from 300 K to 400 K.

## 3. Results & discussion

### 3.1. Sample characterization

The RHEED diffractogram of the $^{57}$FeRh/BaTiO$_3$(100) composite shows a linear structure characteristic of epitaxial layers ( **Fig. 2 A.**), which confirms high-level epitaxy between the alloy and the ceramic.

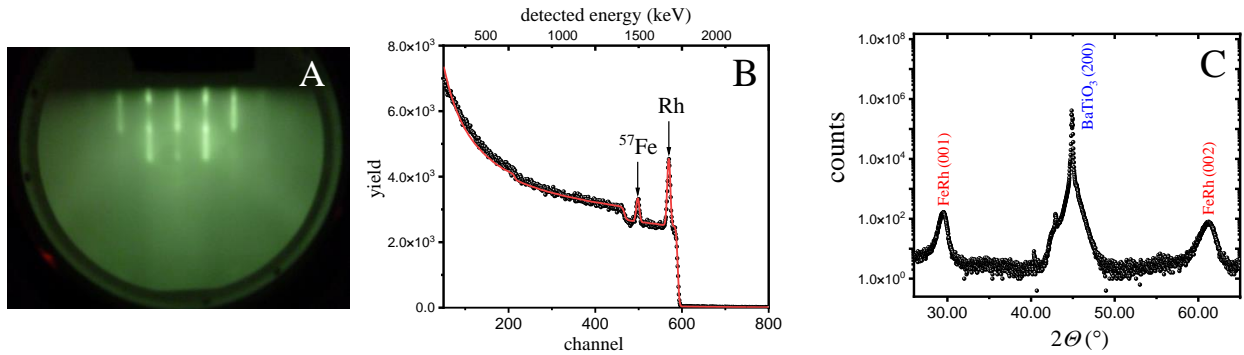

**Fig. 2.** RHEED diffractogram (A); $^4$He$^+$ RBS spectrum at tilt angle of 7°, dots as measured and red line as modelled (B); XRD pattern (C) of the $^{57}$FeRh(B2)/BaTiO$_3$(100) composite

The actual thickness of the $^{57}$FeRh layer and the atomic ratio of the alloy was determined with RBS ( **Fig. 2 B.**) and was found to be (15.7 ± 0.3) nm and $^{57}$Fe$_{0.508 \pm 0.018}$Rh$_{0.492 \pm 0.018}$, respectively.

The XRD pattern shows the B2 type FeRh(001) peak around 30°, the FeRh(002) peak around 61° and the BaTiO$_3$ (200) peak around 45° ( **Fig. 2 C.**). The lattice parameter was estimated to be (3.0361 ± 0.0004) Å, which is close to the value reported for the B2 FeRh phase [61]. Since no peak corresponding to other possible structure of FeRh (e.g. to the paramagnetic A1 phase) was found in the pattern, the alloy can be considered structurally homogeneous. Together, RHEED, RBS and XRD results confirmed the successful, high-quality sample preparation.

### 3.2. Determination of the (sub)layer structure of the FeRh/BTO composite

For the GI-NRS model, the iron microenvironments in the sample were determined by CEMS, from a spectrum ( **Fig. 3 A.**) taken at 294 K. The structural model, outlined in more detail in the **supplementary material**, was the following.

It was presumed that the chemical composition of the FeRh film was uniform and very close to the stoichiometric ratio Fe$_{50}$Rh$_{50}$. In spite of the fact that the phase diagram of the bulk Fe-Rh system [30] predicts pure AFM state for Fe$_{50}$Rh$_{50}$ at 294 K, a coexistence of FM and AFM phases was allowed for as it was previously observed by other authors in thin films [49, 50].

Furthermore, it was allowed for the random presence of both Fe and Rh antisite atoms of which only the Fe atoms contribute to the CEMS spectrum. In next-neighbor approximation, a sextet and a singlet are expected for Fe antisite atoms in the FM and the AFM state, respectively [47, 63]. Using the same approximation, the contribution of Fe atoms in their regular sites can be described [47] by binomial distributions [62] belonging to a certain (pretty small) antisite density. In this respect, four configurations can be distinguished depending on whether in the immediate environment of the selected Fe atom there was a local Fe or Rh excess and whether it was in FM or AFM state.

Indeed, in line with data from previous Mössbauer studies [47, 63, 64], four binomial distributions, a sextet and a singlet proved to be capable of describing the relevant part of the CEMS spectrum. Two and two of the binomial distributions apiece with unperturbed hyperfine fields (*HFF*) of 27.67 T and 25.02 T belonged to the FM and AFM states, respectively [63, 65]. Using four rather than two binomial distributions was made necessary by the fact that both for the FM and the AFM states only two pairs of very different hyperfine perturbation parameters $\Delta\delta$ and $\Delta H$ described the relevant part of the spectrum satisfactorily. The sextet of isomer shift $\delta = 0.34$ mm/s and *HFF* = 39.18 T was typical for Fe antisite atoms in the FM state [63]. The intensity of the singlet belonging to Fe antisite atoms in the AFM phase ($\delta = 0.35$ mm/s) was $(1 \pm 1)$ %, i.e. not really significant.

In addition, a further very broad, unresolved component of 9 % intensity was identified in the CEMS spectrum and phenomenologically fitted with a pseudo-Voigt profile of Lorentzian and Gaussian widths 0.40 mm/s and 3.17 mm/s, respectively. The isomer shift of this anomalous component turned out to be $(0.37 \pm 0.07)$ mm/s, a value typical for high-spin $Fe^{3+}$ atoms like those in oxides. Luckily, such a broad component has absolutely no impact on the GI-NRS quantum-beat patterns as it generates no delayed photons beyond about 1 ns.

The unusual existence of the FM phase at room temperature, a rare occurrence in bulk FeRh but quite common in the case of thin FeRh films is attributed to the small lattice mismatch between the BTO and the FeRh, as it was previously reported [48, 50, 66, 67, 68]. The detailed description of the model used for the deconvolution of the CEMS spectrum can be found in the **supplementary material**.

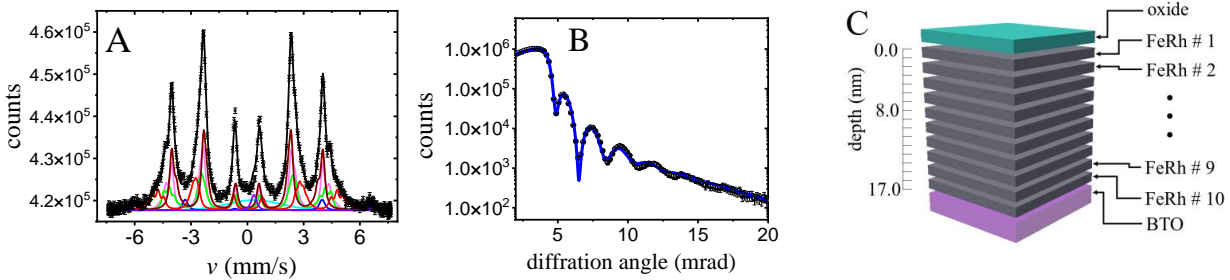

**Fig. 3.** 294 K CEMS spectrum (A), dots as measured, colored lines as simulated, for details check the supplementary material; and X-ray reflectogram, dots as measured, lines as modelled, of the sample (B); and the arrangement of the (sub)layers in the GI-NRS model (C)

The analysis of the XRR reflectogram ( **Fig. 3 B.**) revealed that the actual layer structure of the sample was $Fe_xRh_yO_z$ [(1.4 ±0.2) nm]/$^{57}$FeRh(16.8 ±0.3) nm]/BaTiO$_3$(100), where "$Fe_xRh_yO_z$" is an unspecified iron oxide in which some Fe atoms may be replaced by Rh.

In the GI-NRS model, to be able to model the depth profile of the phase transition, the FeRh layer was divided into ten sublayers, each representing 1/10 of its thickness ( **Fig. 3 C.**). Each of these sublayers was described as a homogeneous composition of all (spectral) components from CEMS results (excluding the oxide component). During the analysis of the GI-NRS quantum-beat patterns, the Mössbauer parameters of the corresponding components were kept at the same values in each sublayer, taking into account the temperature-dependent changes of the parameters. In addition, within the FM and AFM phases, the ratios of the microenvironments were fixed in each sublayer at the ratios derived from CEMS. As a result, the only two fitting parameters which could vary between the individual FeRh sublayers (at any given combination of temperature, electric field and magnetic field were the relative amounts of the FM and AFM FeRh phases $X_{FM}$ and $X_{AFM}$, respectively Since $X_{FM}$ was modeled as 1-$X_{AFM}$ in each sublayer, the only remaining independent parameter was $X_{AFM}$ (layer), which is used in the present article as the only parameter to describe the phase composition of the sublayers.

By simultaneously evaluating the three corresponding GI-NRS quantum-beat patterns apiece that were measured with the same experimental parameters but at different grazing angles; $X_{AFM}$ values were determined individually in each FeRh sublayer at each temperature based on similar principles that were described in Refs [69, 70, 71]. In **Fig. 4.** one can see the calculated intensity (in arbitrary units) in depth of the sample as the function of energy detuning from the resonance (Mössbauer drive velocity). It is apparent that at the resonance peaks the information depth is quite shallow, however one has to keep in mind that in the measured quantum-beat patterns the whole broadened lines rather than only the peaks play role.

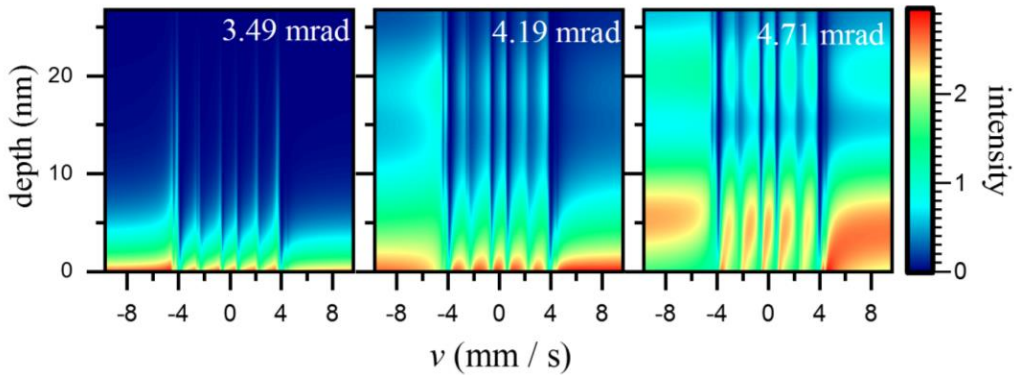

**Fig. 4.** Calculated information depths of the GI-NRS measurement at different grazing angles

The calculated $X_{AFM}$ values, along with their uncertainties, at each temperature (and for each magnetic/electric field combination), can be found in **Tables 3. – 16.** of the **supplementary material**. The fit of the above detailed model to the measured data points, as well as the effects of the different grazing angles, external magnetic and electric fields on the GI-NRS quantum-beat patterns are shown in **Fig. 5.**

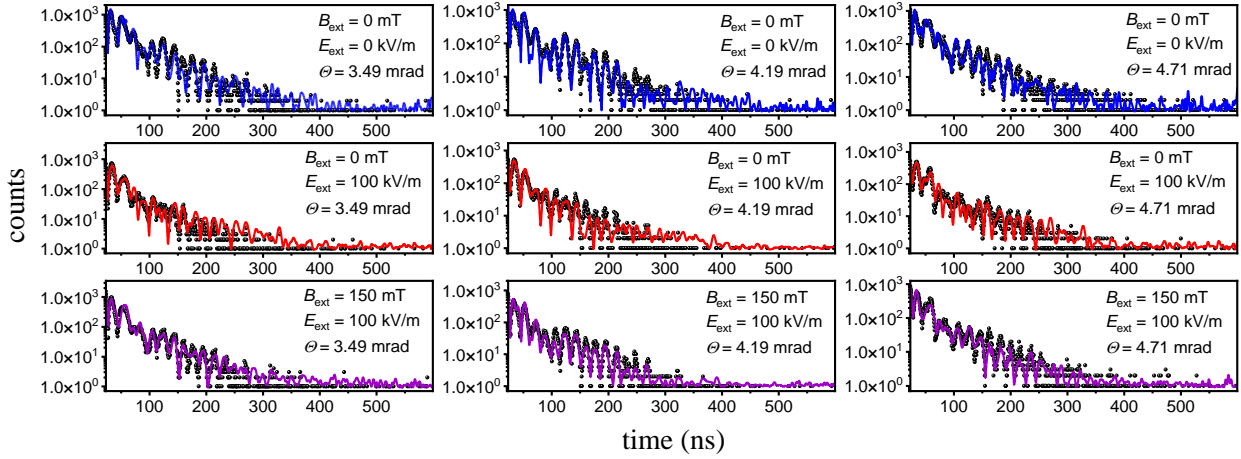

**Fig. 5.** The effect of the different grazing angles and external magnetic/electric fields on the GI-NRS quantum-beat patterns during the cooling phase at 352 K; dots as measured and lines as modelled; the error bars of the measured data points are smaller, than the data marks

### 3.3. Depth-resolved determination of the phase transition of the alloy as a function of temperature, without external magnetic or electric field

The determined $X_{AFM}$ per layer values show that the 1st and the 10th FeRh sublayers contain very little AFM phase, regardless of whether this phase overwhelmingly dominated the $^{57}$FeRh alloy at 299 K without applied magnetic- or electric fields (dark blue bars in **Fig. 6.**). This observation corresponds well to the previous PNR study where similar depth structure was described in comparably thick FeRh films [50].

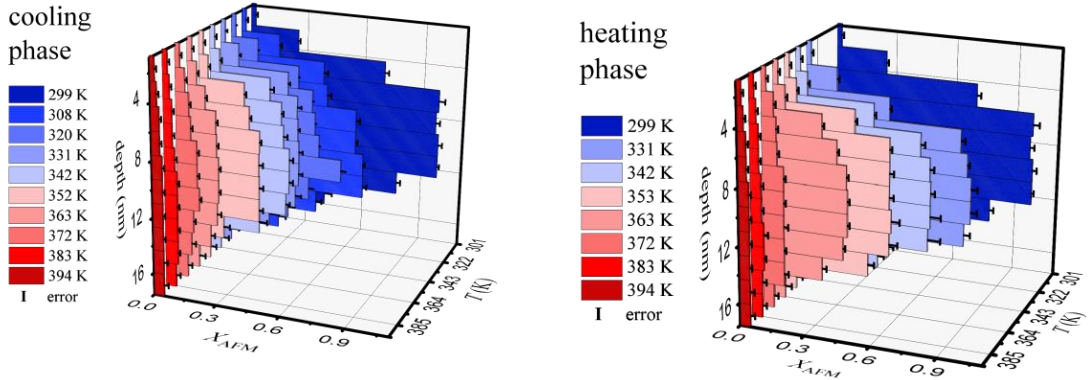

**Fig. 6.** $X_{AFM}$ as function of depth and temperature; without applied electric or magnetic field

The analysis of the further GI-NRS quantum-beat patterns has also revealed that these inherently low $X_{AFM}$ values observed in the 1st and 10th sublayers barely change regardless of temperature of the measurement (**Fig. 6.**). This result indicates that the magnetic structures of these sublayers, at bottom and top of the FeRh layer, are stabilized in FM state due to surface and interface-related effects. In the case of the 10th sublayer, this effect may be explained by the small lattice mismatch between FeRh and BTO inducing an epitaxial strain which favors the FM state, by the same mechanism as described in Refs. [72, 73, 74, 75]. Meanwhile, in the case of the 1st sublayer, oxidation-based processes may be the triggers. These experimental results indicate that the whole FeRh layer can never reach pure AFM phase as an effect of only temperature, which is consistent with previous *ab initio* calculations [76].

In contrast to the top and bottom sublayers, in the middle (~3rd – 8th) sublayers the $X_{\mathrm{AFM}}$ values are dramatically increasing with the decrease of temperature, changing from "fully" FM phase (at 394 K) to a "completely" AFM phase (at 299 K) (**Fig. 6.** and **Fig. 7.**).

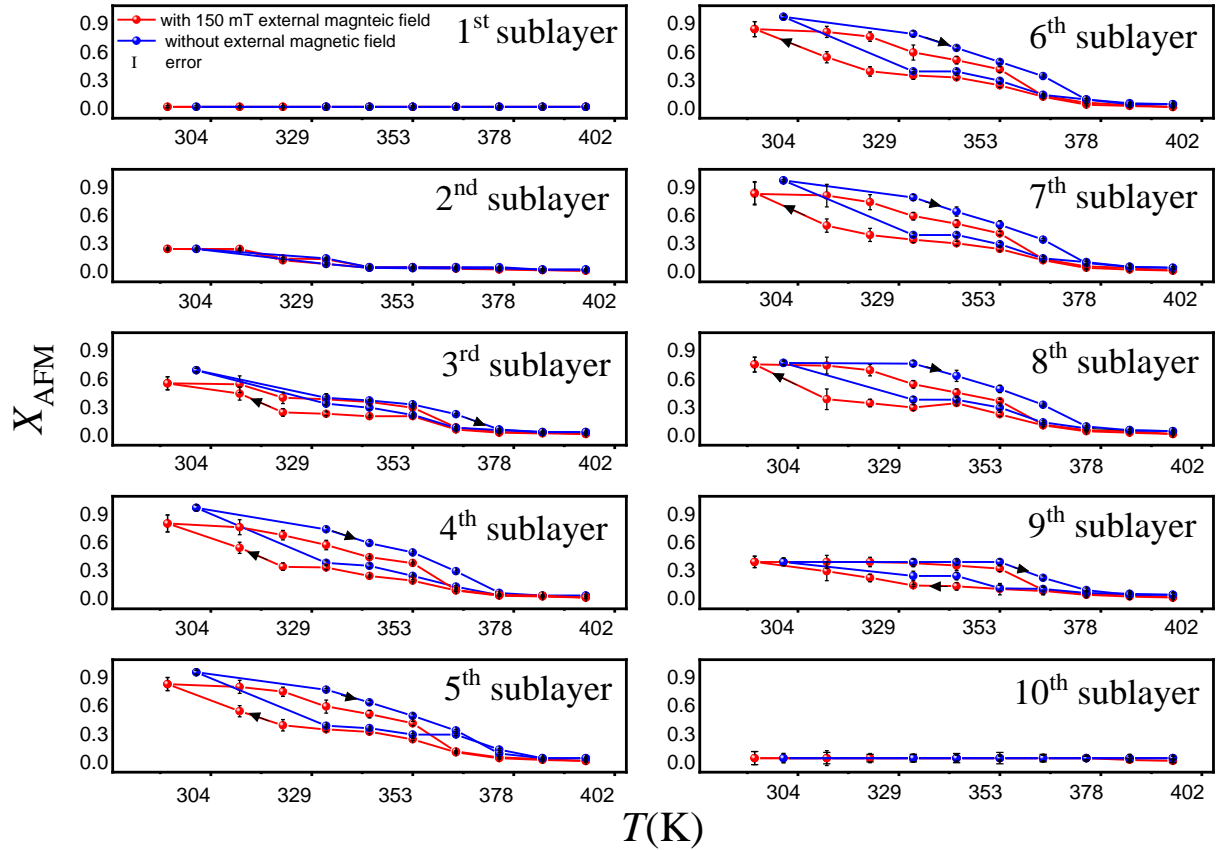

**Fig. 7.** $X_{\mathrm{AFM}}$ as a function of temperature and external magnetic field per sublayer, without external electric field, the black arrows indicate the directions of the temperature changes

As the shapes and slopes of the plotted $X_{\mathrm{AFM}}$ - $T$ curves are similar in each middle sublayer (blue lines in **Fig. 7.**), one can presume that the AFM ↔ FM phase transition occurred (vertically) homogeneously. These findings contradict the results of Ref. [48] where the heating-induced AFM → FM phase transition propagated from top to bottom. Note, however, that the substrate and the thickness of the FeRh layer was different in that study and, therefore, the differences between the results of the present study and those of Ref. [48] may indicate that the depth profile of the phase transition is a function the FeRh layer thickness.

Considering that $X_{AFM}$ as a function of temperature did not change abruptly in any sublayer, it can be presumed that the phase transition of the alloy was also continuous. During mapping of the FeRh surface, similar continuous and homogeneous temperature-induced AFM / FM phase transition, the Volmer-Webber island type growth was also found [77, 78, 79]. The combination of this previous result (horizontally continuous and homogeneous domain growth) with that of the present paper (vertically continuous and homogeneous domain growth) indicates that the phase transition of the alloy was homogeneous and continuous in all three dimensions throughout the layer (except for the top and the bottom sublayers) if it was triggered by temperature change. From a technology viewpoint, these results showed that the alloy layer always has to be at least 4 nm thick in the FeRh/BTO composite if temperature-induced AFM $\leftrightarrow$ FM phase transition is required.

### 3.3.1 Magnetometry control of the temperature-induced phase transition of the alloy

The phase transition described in **3.3.** was strongly supported by magnetometry experiments. Magnetization was calculated using the substrate area (25 mm²) and the FeRh film thickness (16.8 nm) determined by XRR. The coercive force was extracted, after correcting for a ± 0.09 mT sweep-direction dependent offset of the superconducting magnet at every temperature and orientation. It was found that the (010) magnetization of the $^{57}$FeRh/BaTiO$_3$ composite increased from ~400 emu/cm³ to ~700 emu/cm³ as the temperature increased from 300 K to 400 K ( **Fig. 8 A.**) that corresponds well to the temperature-induced AFM $\rightarrow$ FM phase transition. The existence of multiple ferromagnetic sublayers (top and bottom sublayers in the GI-NRS model) was also supported by magnetometry, as the hysteresis loop(s) at lower temperatures consisted of two separated FM components ( **Fig. 8 A.**), one with lower, and one with higher coercive force (**Table 2.** of the **supplementary material**).

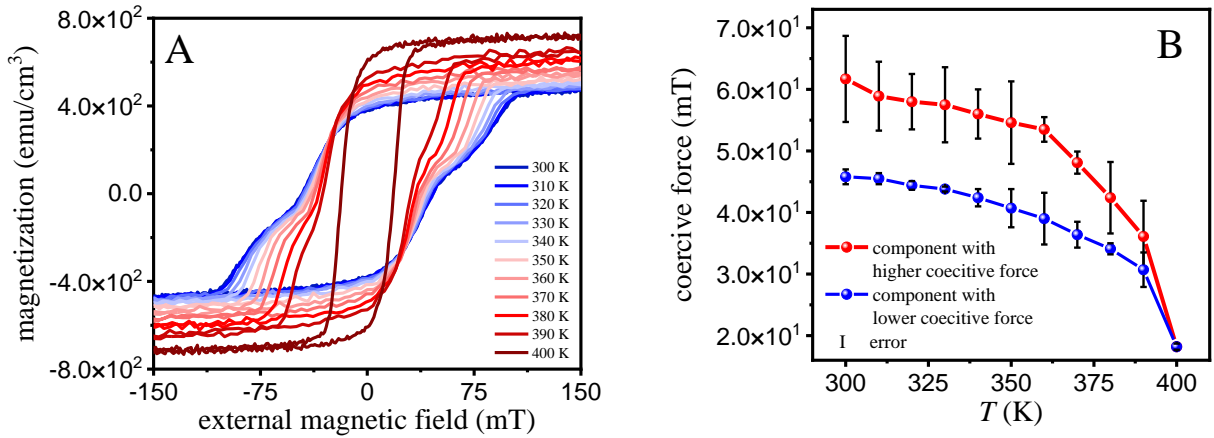

**Fig. 8.** Magnetization at (010) direction (A) and coercive forces (B) as a function of magnetic field and temperature in the FeRh/BTO composite

As the sample was heated up, the coercivity separation decreased ( **Fig. 8 B.**), which also corresponds well to the FM / AFM / FM $\rightarrow$ FM transition as determined by GI-NRS. The abrupt decrease in the coercive force(s) around 400 K can be attributed to the tetragonal $\rightarrow$ cubic phase transformation of the BTO which forces the coupled FeRh layer into FM phase due to epitaxial strain. At last, the results of magnetometry also show that 150 mT external magnetic field used in the GI-NRS experiments was enough to reach the full magnetic saturation.

### 3.4. Depth-resolved determination of the phase transition of the alloy as a function of temperature in external magnetic field

The 2nd group of GI-NRS measurements showed that, regardless of the applied external magnetic field, the 1st and 10th sublayers always remained completely in FM phase (**Fig. 9.**). Therefore, one can assume that in this composite there exist no temperature which can result in a pure AFM phase regardless of whether external magnetic field is applied or not.

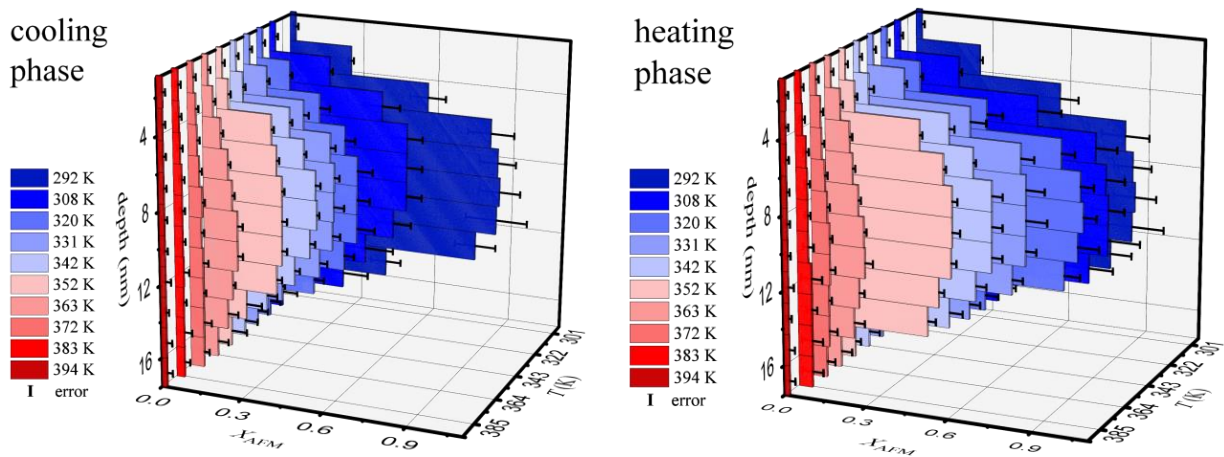

**Fig. 9.** $X_{AFM}$ as function of depth and temperature, with 150 mT external magnetic field applied

The similar shapes and slopes of the $X_{AFM} - T$ curves from the first two GI-NRS measurement series (**Fig. 7.**) indicate that the AFM ↔ FM phase transitions occurred by the same mechanism in both measurement series. Therefore, it can be assumed that the application of external magnetic field had no localized effect on this composite but it homogeneously preserved the quantity of the FM phase during the cooling phase. However, the comparison of the $X_{AFM}$ values also revealed, that the AFM phase ratio in the middle sublayers differed depending on the strength of the magnetic field applied during the cooling period (**Fig. 7.**) similar to what was previously described for the 'average AFM phase ratio' [80]. The difference between the zero-field-cooled and in-field-cooled measurements indicated that ferromagnetic clusters were always present in the middle FeRh sublayers [81] supporting the previously hypothesized homogeneous domain growth mechanism.

### 3.5. Depth-resolved determination of the phase transition of the alloy as a function of temperature and electric field in the FeRh/BTO multiferroic

The $X_{AFM}$ values measured with applied electric field demonstrated that, in contrast to the effects of temperature, the voltage could even transform the bottom FM sublayer into AFM phase in this multiferroic (**Fig. 10.**). This different behavior was a consequence of the mechanical strain dependence of the $X_{AFM}$ phase, which originates from the combination of the epitaxial coupling between the FeRh and BTO layers and the electric field induced piezoelectric strain of the BTO [82].

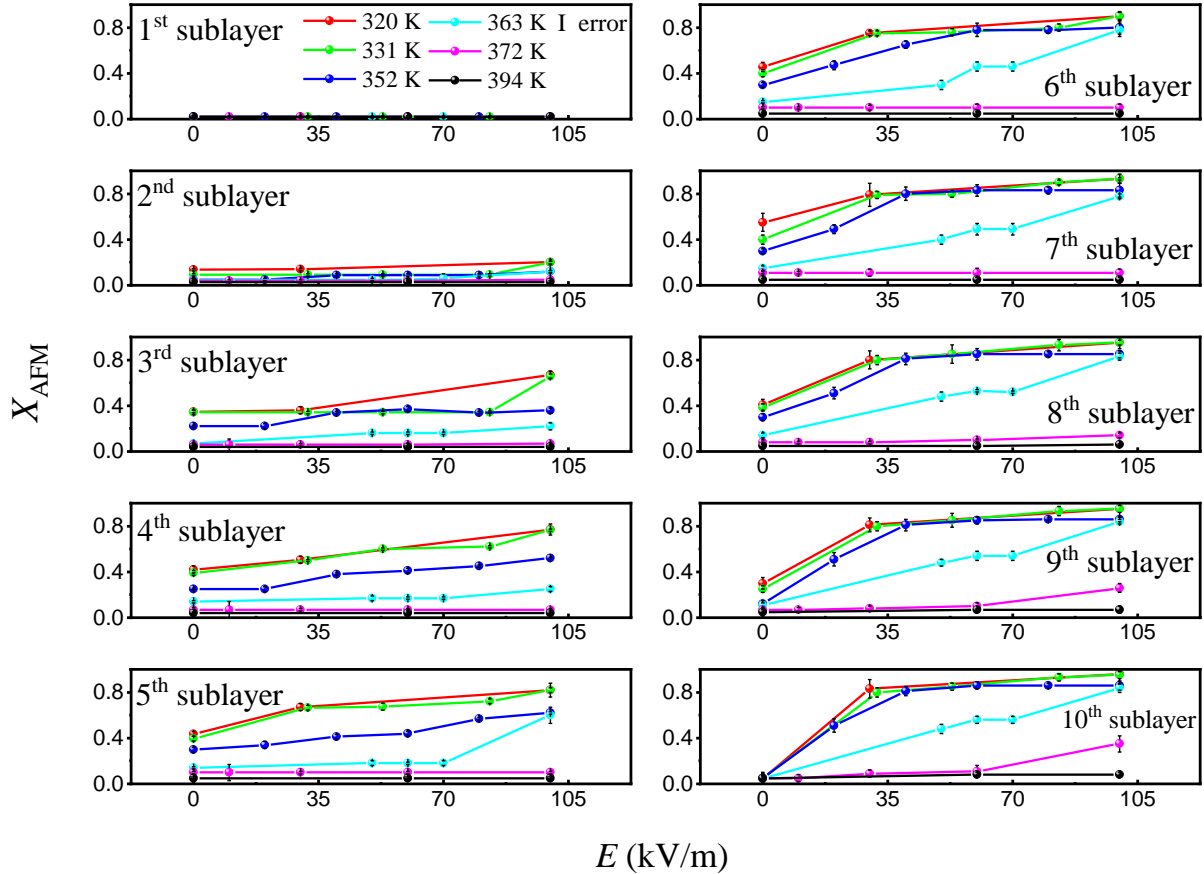

**Fig. 10.** $X_{AFM}$ as a function of temperature and external electric field per layer, without applied magnetic field

The comparison of the $X_{AFM} - E$ curves measured at different temperatures (**Fig. 10.**) revealed that below 320 K and above 372 K the external electric field had no measurable effect on the AFM / FM phase ratio of the alloy. Below 320 K there was simply not enough FM phase in the alloy for its transformation to be significant. In contrast, above 372 K the FM phase was too stable to be significantly transformed to AFM phase by the epitaxial coupling. However, within this temperature range, the electric-field-induced phase transition could easily overpower the effect of temperature in the ~ 9 nm vicinity of the BTO substrate (**Fig. 10.**). Therefore, this multiferroic coupling is optimal for anticipated devices that can operate with FeRh layer up to 9 nm thickness in this temperature range (optimally at 342 K).

In addition, the plotting of $X_{AFM}$ as a function of depth and $E$ (**Fig. 11.**) demonstrates that the electric field induced FM → AFM phase transition occurred practically completely as soon as $E$ reached a certain limit between 10 kV/m and 30 kV/m. However, once the electric field reached the threshold, its further increase had only a small effect on the phase ratio of the alloy. The exact range of electric field strength that triggers the phase transition is determined by the temperature.

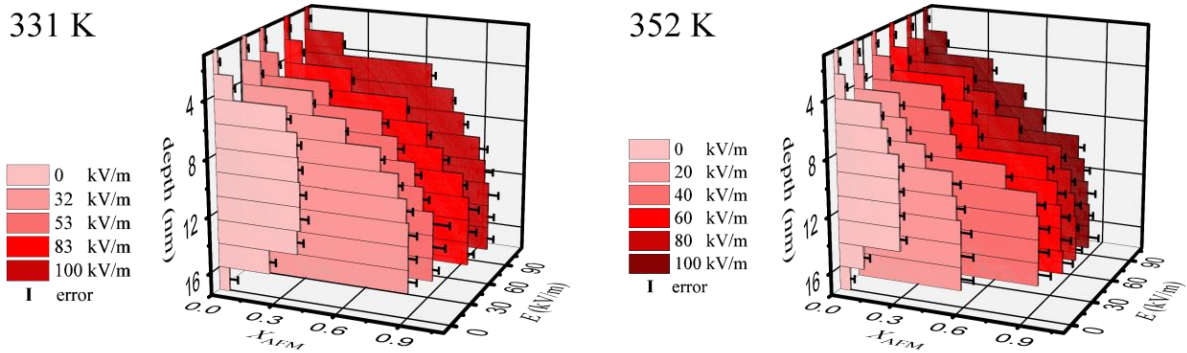

**Fig. 11.** $X_{AFM}$ as function of depth and electric field, without external magnetic field; during electric field buildup

Admittedly, there is also a downside to this electric-field effect: contrary to the temperature-induced transition, this phenomenon is not automatically reversible; i.e. $X_{AFM}$ does not convert back completely when the electric field drops back to zero. Conversely, the initial phase ratio of the alloy can be completely restored by heating and then cooling back.

### 3.6. Comparison of the depth-dependent effects of temperature, external magnetic and electric fields on the phase transition of the alloy

The results of the 4[th] GI-NRS experiments showed that the presence of an external magnetic field reduced the effect of the electric field at all temperatures where it had a measurable effect, not at the FeRh/BTO interface, though (**Fig. 12.**). This inhibition of AFM phase formation is very similar to the effect described during the cooling phase in **3.4.** Therefore, the four GI-NRS experiments jointly revealed that the effects of temperature, external magnetic and electric fields on the phase ratio of the alloy in the FeRh/BTO multiferroic may amplify or even eliminate each other (**Fig. 12.**). However, the relative strength of these effects was not homogeneous throughout the FeRh layer; in the ~4 nm vicinity of the substrate the effect of the electric field was vastly stronger than the effects of either external magnetic field or temperature (within the given experimental conditions). Further away from the interface, the AFM / FM phase ratio was determined by the balance of temperature, magnetic and electric field. In addition, all effects of the external electric field could be eliminated by the external magnetic field up to a certain height, measured from the top of the substrate that was determined by the temperature. The cancellation effect extended to ~12 nm above the substrate at 352 K; but only up to ~9 nm if the sample was cooled down to 331 K (**Fig. 12.**). Therefore, the FeRh thickness where the external electric field could trigger the FM → AFM phase transition, was determined by the combination of temperature and the strength of external magnetic field.

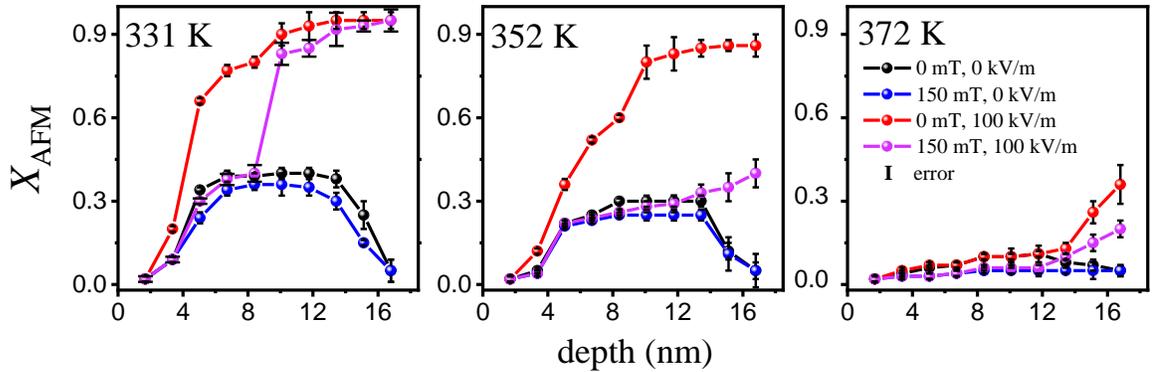

**Fig. 12.** $X_{AFM}$ as a function of temperature, external electric and magnetic fields in the FeRh/BTO composite; during cooling phases

## 4. Conclusion

In summary, details of depth-dependent effects of temperature, external magnetic and electric fields on the AFM / FM phase ratio of the alloy in the FeRh/BTO multiferroic were determined by GI-NRS measurements.

It was established that the FM phase of the top and bottom part of the FeRh layer can never be transformed into the AFM phase by temperature alone, regardless of whether magnetic field is present or not. Beyond these regions, the decrease of temperature induces a continuous and homogeneous growth of AFM domains in the FeRh alloy, regardless of the strength of the external magnetic field. However, the maximum achievable AFM fraction and the temperature of the main phase transition depend on the external magnetic field.

In contrast, the applied electric field induces the magnetic phase transition by a completely different mechanism in the FeRh/BaTiO$_3$ composite: it triggers a bottom → up oriented mechanical strain that enforces the abrupt FM → AFM phase transition once the strength of the electric field reaches a certain limit. By this mechanism, the external electric field is capable of transforming even the lowest FM sublayer into the AFM phase but, at the same time, it also limits the layer thickness where the voltage can change the phase ratio of the alloy. However, this thickness limit is adjustable by changing the temperature and external magnetic field.

From an engineering point of view, the results of the present study indicate that for the optimal utilization of the multiferroic coupling of the FeRh/BTO multiferroic, the thickness of the alloy should be between 4 and 12 nm; and the devices based on this coupling should operate around 342 K.

## CRediT authorship contribution statement

**Attila Lengyel**: Investigation, Data curation, Methodology, Writing - Original Draft. **Gábor Bazsó**: Investigation. **Aleksandr I. Chumakov**: Investigation, Resources. **Dénes L. Nagy**: Investigation, Methodology, Writing - Review & Editing. **Gergő Hegedűs**: Resources. **Dimitrios Bessas**: Resources. **Zsolt E. Horváth**: Investigation. **Norbert M. Nemes**: Investigation. **Maria

**A. Gracheva**: Investigation, Writing - Review & Editing. **Edit Szilágyi**: Investigation, Review & Editing. **Szilárd Sajti**: Calculation, Software. **Dániel G. Merkel**: Conceptualization, Supervision, Investigation, Data curation, Methodology, Funding acquisition.

## Declaration of Competing Interest

The authors declare that they have no known competing financial interests or personal relationships that could have appeared to influence the work reported in this paper.

## Data availability

The data that indicated the above-described results are available in the CONCORDA repository under entry: (the link will come as soon as the article is accepted).

## Acknowledgements

We acknowledge the European Synchrotron Radiation Facility for provision of synchrotron radiation resources at the beamline ID18.

This research did not receive any specific grant from funding agencies in the public, commercial, or not-for-profit sectors.

## Appendix A. Supplementary material

Supplementary data associated with this article can be found in the online version at (the link will come as soon as the supplementary material gets it).